\documentclass[aps,pra,showpacs,preprint]{revtex4-1}
\usepackage{amsmath}
\usepackage{dcolumn}
\usepackage{graphicx}
\usepackage{bm}
\usepackage{textcomp}
\usepackage{upgreek}
\usepackage{enumerate}
\usepackage{epstopdf}

\def\bra#1{\mathinner{\langle{#1}|}}
\def\ket#1{\mathinner{|{#1}\rangle}}

\newcommand{\etal}{{\it et al.\ }}

\begin{document}
	\title{Tuning of the Hanle effect from EIT to EIA using spatially separated probe and control beams}
	
	\author{Mangesh Bhattarai}
	\author{Vineet Bharti}
	\author{Vasant Natarajan}
	\email{vasant@physics.iisc.ernet.in}
	
	\affiliation{Department of Physics, Indian Institute of Science, Bangalore-560012, India}

%
%
%
%
%
%

\begin{abstract}
We demonstrate a technique for continuous tuning of the Hanle effect from electromagnetically induced transparency (EIT) to electromagnetically induced absorption (EIA) by changing the polarization ellipticity of a control beam. In contrast to previous work in this field, we use spatially separated probe and control beams. The experiments are done using magnetic sublevels of the $ F_g = 4 \rightarrow F_e = 5 $ closed hyperfine transition in the 852 nm D$_2$ line of $^{133}$Cs. The atoms are contained in a room temperature vapor cell with anti-relaxation (paraffin) coating on the walls. The paraffin coating is necessary for the atomic coherence to be transported between the beams. The experimental results are supported by a density-matrix analysis of the system, which also explains the observed amplitude and zero-crossing of the resonances. Such continuous tuning of the sign of a resonance has important applications in quantum memory and other precision measurements.
\end{abstract}

	\flushbottom
	\maketitle
	\thispagestyle{empty}

\section{Introduction}
Interaction of light fields with multilevel atoms can result in long lived atomic coherences, which has important applications in diverse areas such as next-generation atomic clocks \cite{VAN05}, sensitive magnetometry \cite{BGK02}, precision measurements \cite{KPW05}, and optical memories \cite{LST09}. These experiments are usually done in vapor cells, where the presence of a buffer gas or anti-relaxation (paraffin) coating on the walls increases the coherence time and results in a narrower linewidth \cite{BNW97,BOB66,RBA16}.

Atomic coherences have also been studied in the Hanle configuration by measuring the transmission of a probe beam as a function of magnetic field. The resulting dark resonance is called electromagnetically induced transparency (EIT), while the bright resonance is called electromagnetically induced absorption (EIA). Both EIT and EIA have been studied in the past \cite{RMW97,RZV01}. More recently, tuning from EIT and EIA has been demonstrated in $^{87}$Rb \cite{RAP10}, $^{39}$K \cite{GFL17}, and $^{133}$Cs \cite{RBB17}. The present work also demonstrates tuning from EIT to EIA, but is different because it uses spatially separated control and probe beams.

We further show that the experimental results are supported by a detailed density-matrix analysis of the system. The presence of spatially separated beams requires the use of a theoretical model with three regions. And the coherence survives between the two beams only when there is paraffin coating on the walls. Such spatial transport of atomic coherences is unique, and has important applications in transporting light \cite{ZMK02}, slow-light beam splitter \cite{XKH08} and anti-parity time symmetry \cite{PCS16}. The theoretical model presented in this work can also be used to provide a theoretical analysis of these experiments.

\section{Experimental details}
\begin{figure*}
	\centering
	\includegraphics[width=.9\textwidth]{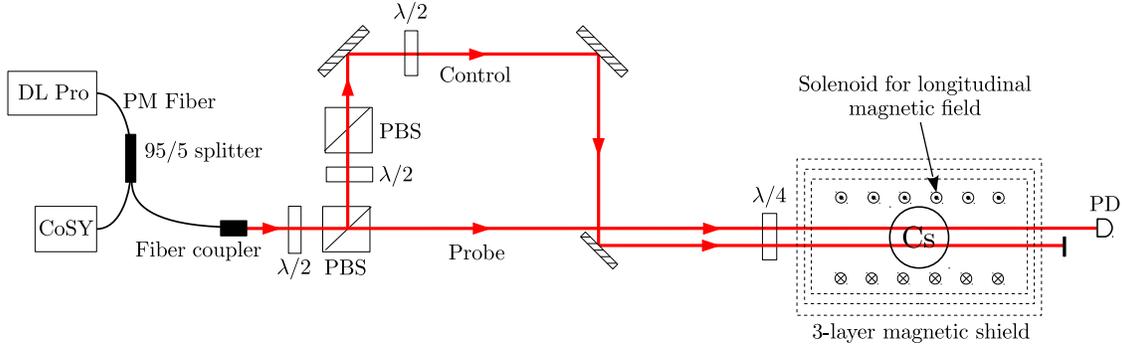}
	\caption{Schematic of the experiment. Figure key: $\lambda$/2 -- half-wave retardation plate; $\lambda$/4 -- quarter-wave retardation plate; PM fiber -- polarization-maintaining fiber, PBS -- polarizing beam splitter cube; PD -- photodiode. }
	\label{expsetup}
\end{figure*}

A schematic diagram of the experimental setup is given in Fig.~\ref{expsetup}. The light is derived from a Toptica DL Pro laser operating near the 852 nm D$_2$ line of $^{133}$Cs. The light from the laser is coupled into a polarization maintaining fiber. The light passes through a 95/5 splitter; after which 5\% of it passes into a CoSY system for saturated absorption spectroscopy, and 95\% of it is coupled to free space through a fiber coupler. The output beam after the coupler  has a Gaussian intensity distribution with $1/e^2$ diameter of 3 mm. The beam is passed through a half-wave ($\lambda$/2) retardation plate and polarizing beam splitter cube (PBS) combination, which splits the beam into the probe and control beams. The angle of the retardation plate is adjusted to control the probe power $P_p$. The control beam power $P_c$ is adjusted using another $\lambda$/2 retardation plate and PBS combination. 

The two beams go into a spherical glass vapor cell containing $^{133}$Cs atoms at room temperature. The sphere has diameter of 75 mm. It has paraffin coating on the walls, which as mentioned before increases the coherence time among the magnetic sublevels of the ground state. The cell is placed within a solenoid of length 640 mm and diameter 190 mm, containing 1800 turns of 0.35 mm diameter wire. The solenoid is used to apply the required longitudinal magnetic field, ranging up to a few mG.

The vapor cell and the solenoid are kept inside a 3-layer \textmu-metal magnetic shield, which reduces stray external fields to less than 1 mG. A small residual transverse field of the order of 0.1 mG is present inside the shield. The probe beam after the shield is detected on a photodiode. The photodiode signal---which is proportional to probe transmission---is collected on the input channel of a data acquisition card. An output channel of the same card is used to sweep the current in the field-producing solenoid. 

The probe and control beams are parallel to each other and separated by distance of 10 mm. The probe beam is always circularly polarized with one handedness, which is verified by adjusting the angle of the quarter-wave ($\lambda$/4) retardation plate (before the cell) for maximum reflection through a PBS. For the same angle of the $\lambda$/4 plate, the control beam will have orthogonal circular polarization because it is the reflected beam from a PBS. The ellipticity of the control beam $\epsilon_c$ is varied by changing the angle of the $\lambda$/2 retardation plate in its path, which changes the direction of linear polarization entering the $\lambda$/4 retardation plate.

\section{Theoretical analysis}

\subsection{Density-matrix model}

\begin{figure}[ht]
	\centering
	\includegraphics[width=.3 \textwidth]{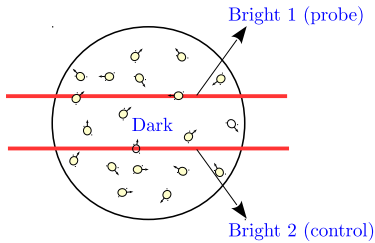}
	\caption{Spatially separated probe and control beams used in the experiment. The regions are labeled Bright 1, Dark, and Bright 2. Atomic coherences are transported by thermal motion in the vapor cell, which survives only if there is paraffin coating on the walls.}
	\label{regionsfortheory}
\end{figure}

In this subsection, we develop a three-region density-matrix model to analyze the spatially separated beam arrangement. Our model is based upon the multi-region model given by S. Rochester \cite{ROC10}. The region of interaction for atoms is divided into three regions labeled as dark, bright 1 (probe), and bright 2 (control) as shown in Fig.~\ref{regionsfortheory}. In the dark region only magnetic field is present, whereas in the bright regions both light and magnetic field are present. 

The probe and control beams are given by the real parts of following expressions \cite{ABR10}
\begin{equation}
\begin{aligned}
\mathcal{E}_p &= \mathcal{E}_{p0} \left\{ \cos\left(\epsilon_p + \dfrac{\pi}{4}\right) {\hat{\epsilon}_{-1}}- \cos\left(\epsilon_p - \dfrac{\pi}{4}\right) {\hat{\epsilon}_{+1}}\right\} e^{-i\omega_l t} \\
\mathcal{E}_c &= \mathcal{E}_{c0} \left\{ \cos\left(\epsilon_c + \dfrac{\pi}{4}\right) {\hat{\epsilon}_{-1}}- \cos\left(\epsilon_c - \dfrac{\pi}{4}\right) {\hat{\epsilon}_{+1}}\right\}e^{-i\omega_l t}
\end{aligned}
\end{equation} 
Here, $\mathcal{E}_{p0(c0)}$ and $\epsilon_{p(c)}$ are the respective amplitude and ellipticity of probe (control) beam, and $\hat{\epsilon}_{\mp 1}$ are the unit vectors along $\sigma^{\mp}$ directions. 

\begin{figure}[ht]
	\centering
	\includegraphics[width=.6 \textwidth]{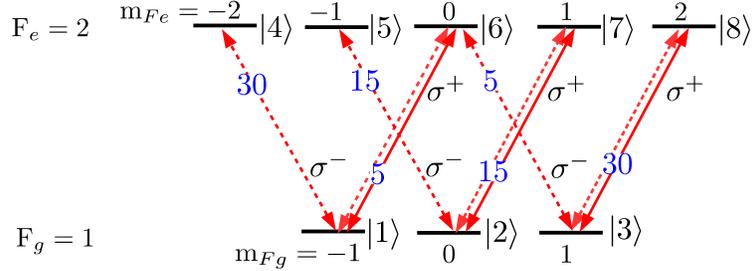}
	\caption{Magnetic sublevels for $F_g =1 \rightarrow F_e =2$  transition. The probe beam has $ \sigma^+ $ polarization, and transitions coupled by it are shown with solid lines. The control beam polarization is varied from $ \sigma^+ $ to $ \sigma^- $, with transitions shown by dotted lines. The numbers on each arrow are the normalized Clebsch-Gordan coefficients (relative strength) of each transition.}
	\label{levels1to2}
\end{figure}

We present theoretical results for an $F_g = 1  \rightarrow F_e = 2$ transition, which is the simplest case for $F_g  \rightarrow F_e = F_g+1$ closed transitions. The magnetic sublevels for the $F_g =1 \rightarrow F_e =2$ transition, labeled $\ket 1$ to $\ket 8$, are shown in Fig.\ \ref{levels1to2}. The probe field transitions are shown with solid lines and control field transitions are shown with dashed lines. We consider that the probe beam has $\sigma^+$ polarization and the control beam polarization is varied from  $\sigma^+$ to $\sigma^-$ by changing its ellipticity. 

The time evolution of the system for three regions is given by following set of coupled density matrix equations:
\begin{equation}
\begin{aligned}
\dot{\rho}^{\rm{bright 1}}&=\dfrac{-i}{\hbar}\left[(H_{\rm int}+H_B), \rho^{\rm{bright 1}}\right]-\dfrac{1}{2}\left\{\hat{\Gamma},\rho^{\rm bright 1} \right\}-\gamma_{t1}\rho^{\rm{bright 1}}+\Lambda^{\rm{bright 1}}+\gamma_{t1}\rho^{\rm{dark}} \\
\dot{\rho}^{\rm dark}&=\dfrac{-i}{\hbar}\left[H_B, \rho^{\rm{dark}}\right]-\dfrac{1}{2}\left\{\hat{\Gamma},\rho^{\rm{dark}} \right\}+\dfrac{\gamma_{t2}}{2}(\rho^{\rm{bright 1}} +\rho^{\rm{bright 2} })+\Lambda^{\rm{dark}} -(\gamma_{t2}+\gamma_w)\rho^{\rm{dark}} \\ & \, \, \, \, \, \, + \dfrac{\delta^{\rm gs}}{2F_g+1}\gamma_w \\
\dot{\rho}^{\rm{bright 2}}&=\dfrac{-i}{\hbar}\left[(H_{\rm int}+H_B), \rho^{\rm{bright 2}}\right]-\dfrac{1}{2}\left\{\hat{\Gamma},\rho^{\rm{bright 2}} \right\}-\gamma_{t3}\rho^{\rm{bright 2}}+\Lambda^{\rm{bright 2}}+\gamma_{t3}\rho^{\rm{dark}}
\label{threeregionequations}
\end{aligned}
\end{equation} 
where $H^{\rm bright1}_{\rm int}$ and $H^{\rm bright2}_{\rm int}$ are the Hamiltonian for light-atom interaction in bright 1 and bright 2 regions; $\hat{\Gamma}$ is the relaxation matrix for decay of atoms from excited levels due to spontaneous emission; $\Gamma_e$ is the decay rate of the excited state; $\Lambda^{\rm bright1,bright2,dark}$ represent repopulations of atoms in the ground levels due to decay from excited levels in bright 1, bright 2, and dark regions; $\gamma_{w}$ is ground state relaxation rate due to wall collisions; $\gamma_{t1}$ is the transit-time relaxation rate from  bright 1 to the dark region; $\gamma_{t2}$ is the transit-time relaxation rate from the dark region to the bright regions; $\gamma_{t3}$ is the transit-time relaxation rate from bright 2 to the dark region; and $\delta^{\rm gs}$ is the identity matrix for ground state. 

The repopulation terms ($\Lambda^{\rm bright1,bright2,dark}$) are incorporated by using corresponding branching ratios from excited to ground states. The ground state coherence decay rate ($\gamma_w$) is taken to be consistent with the linewidth of chopped NMOR signal obtained using same cell by Ravi \etal \cite{RBA16}.  The transit-time relaxation rates in the bright regions ($\gamma_{t1}$ and $\gamma_{t3}$) are estimated from the time it takes for a typical Cs atom to traverse the laser beam. The size of the laser beam is taken as the $1/e^2$ diameter. The typical atomic velocity is taken as the rms velocity at room temperature of 300 K. The actual value of room temperature varies by a few degrees from one room to another, but this will not change the value of the rms velocity significantly, which is taken to be 195.6 m/s. The relaxation rate to go out of the dark region ($\gamma_{t2}$) is taken to be proportional to $\gamma_{t1}$, with the proportionality ratio given by the ratio of the volumes of the dark and bright regions inside the cell (about $400$).

The light-atom interaction Hamiltonians for the bright 1 and bright 2 regions are: 
\begin{equation}
\begin{aligned}
H^{\rm bright 1}_{\rm int}&=\dfrac{\hbar}{2} \sum_{i=1}^{3} \sum_{j=4}^{8} \ket{i} \bra{j} \, \Omega^p_{ij} \left[\cos\left(\epsilon_p+\dfrac{\pi}{4}\right) \delta_{i+3,j} - \cos\left(\epsilon_p-\dfrac{\pi}{4}\right) \delta_{i+5,j}\right]+{\rm h.c.}
\label{bright13region}
\end{aligned}
\end{equation} 
\begin{equation}
\begin{aligned}
H^{\rm bright 2}_{\rm int}&=\dfrac{\hbar}{2} \sum_{i=1}^{3} \sum_{j=4}^{8} \ket{i} \bra{j} \, \Omega^c_{ij} \left[\cos\left(\epsilon_c+\dfrac{\pi}{4}\right) \delta_{i+3,j} - \cos\left(\epsilon_c-\dfrac{\pi}{4}\right) \delta_{i+5,j}\right]+{\rm h.c.}
\label{bright23region}
\end{aligned}
\end{equation} 
and the magnetic field Hamiltonian can be described as: 
\begin{equation}
\begin{aligned}
H_{B}&=\mu_B B_z 
\begin{bmatrix}
g_F \sigma^{(z)}_{3 \times 3}    & 0\\
0    &  g_{F'} \sigma^{(z)}_{5 \times 5} 
\end{bmatrix}  +\mu_B B_x 
\begin{bmatrix}
g_F \sigma^{(x)}_{3 \times 3}    & 0\\
0    &  g_{F'} \sigma^{(x)}_{5 \times 5} 
\end{bmatrix}    
\label{BfiledHamiltonian}
\end{aligned}
\end{equation} 
where $\Omega^{p(c)}_{ij}$ is the probe (control) Rabi frequency for the $ij$ $^{\rm th}$ element, $B_z$ is the longitudinal component of the magnetic field, $B_x$ is the residual component of the transverse field, $g_F$ is the Land\`e g-factor of the ground level, $g_{F'}$ is the Land\`e g-factor of the excited level,  $\mu_B$ is Bohr magneton, and $\sigma^{(x)}_{2i+1 \times 2i+1}$ and $\sigma^{(z)}_{2i+1 \times 2i+1}$ are the $x$ and $z$ components of Pauli matrices for spin $i$ particles.

\subsection{Population redistribution due to optical pumping}

\begin{figure}[ht]
	\centering
	\includegraphics[width=.9 \textwidth]{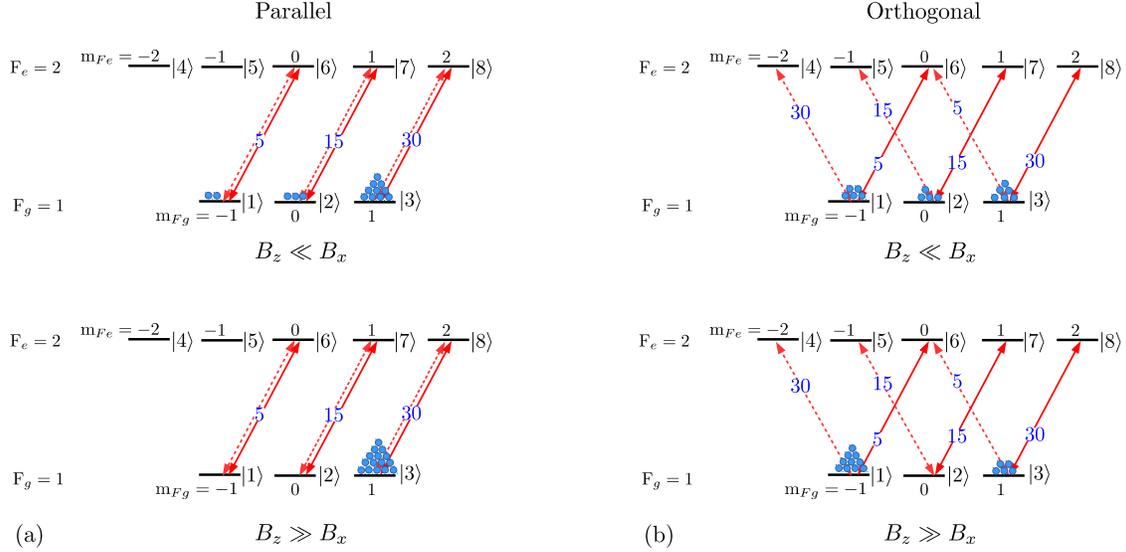}
	\caption{Population redistribution due to optical pumping. (a) Probe and control beams with parallel circular polarizations. Upper part for $B_z \ll B_x$ (no applied field) and lower part for $B_z \gg B_x $ (large applied field). (b) Probe and control beams with orthogonal circular polarizations. Upper part for $B_z \ll B_x $ (no applied field) and lower part for $B_z \gg B_x $ (large applied field).}
	\label{levels_for_explanation}
\end{figure}

The physics behind the detailed theoretical model presented above can be understood by considering population redistribution among the magnetic sublevels due to optical pumping by the probe and control beams. The probe beam is taken to be circularly polarized, as in the experiment. However, optical pumping due to the control beam can be easily understood only when it has circular polarization, while in the experiment it is actually changed from one handedness to another. The results are shown in Fig.~\ref{levels_for_explanation}, with part (a) showing the populations for parallel circular polarizations, and part (b) showing the populations for orthogonal circular polarizations.

For parallel polarizations of the two fields, when no longitudinal field is applied and in the presence of a residual transverse field ($B_z \ll B_x $), most of the population is in the $\ket{3}$ state with less in the $\ket{1}$ and $\ket{2}$ states. The populations are not equal because the quantization axis and the direction of the magnetic field are different. In the presence of a large applied longitudinal field ($B_z\gg B_x $), the quantization axis and the direction of the field are the same. As a result, all the population ends up in the extreme sublevel ($\ket{3}$ state). We can see from Fig.~\ref{levels_for_explanation} that the sum of the net population weighted with the strength of respective transitions coupled by the $\sigma^+$ probe is higher now. This will result in increased absorption as $B_z$ is increased.

When the polarizations are orthogonal, population redistribution for $B_z \ll B_x$ is as shown in upper part of Fig.~4(b). Populations are unequal in the three states, both because of the quantization axis being different from the direction of the field, and the fact that the probe and control powers are different. For the case of $B_z \gg B_x$, both extreme sublevels (states $\ket{1} $ and $ \ket{3}$) have some population because the probe and control are non-zero but unequal. This will result in decreased absorption as $B_z$ is increased.

\section{Results and discussion}

\subsection{Transformation from EIT to EIA}
\begin{figure*}[ht]
	\centering
	\includegraphics[width=.9 \textwidth]{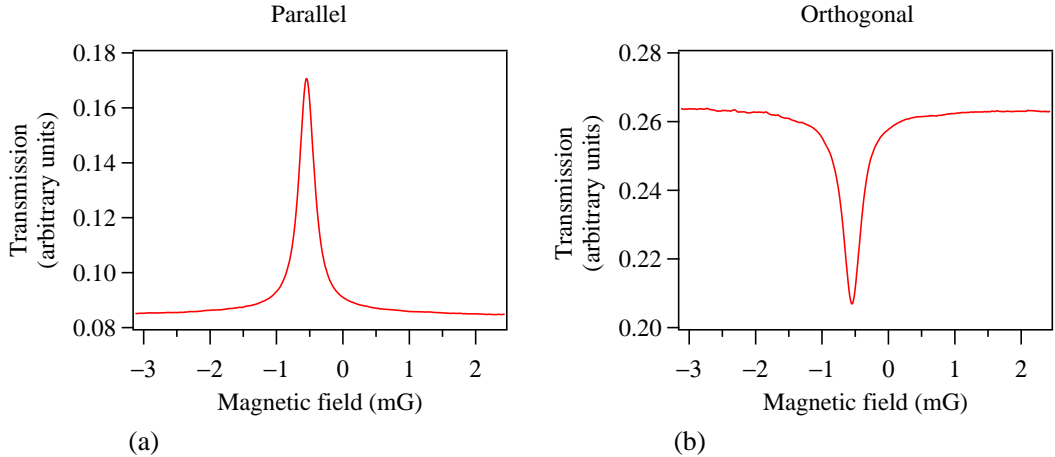}
	\caption{Probe transmission verses longitudinal B field. (a) Hanle EIT when probe and control beams have parallel circular polarizations. (b) Hanle EIA when probe and control beams have orthogonal circular polarizations.}
	\label{plot_45-45}
\end{figure*}

We first consider the transformation from EIT to EIA as the polarizations of the probe and control beams are changed from parallel circular to orthogonal circular. The results, shown in Fig.~\ref{plot_45-45}, shows probe transmission as a function of applied longitudinal B field. The probe beam has a power of $30$ \textmu W, and its ellipticity is fixed at $+45^\circ$. The control beam has a power of 80 \textmu W, and its ellipticity is changed from $+45^\circ$ (parallel to the probe) to $-45^\circ$ (orthogonal to the probe). The former shows Hanle EIT (part (a) of the figure) while the latter shows Hanle EIA (part (b) of the figure). The curves are obtained by averaging for 10 values of probe transmission for each value of B field---this ensures that the system reaches ``steady-state'' in terms of the Larmor precession frequency. 10 such curves are averaged over to account for point-to-point variation in the transmission maximum. The resonances are not centered at 0 field because of the presence of a small residual field inside the shield. 

\begin{figure*}[ht]
	\centering
	\includegraphics[width=.9 \textwidth]{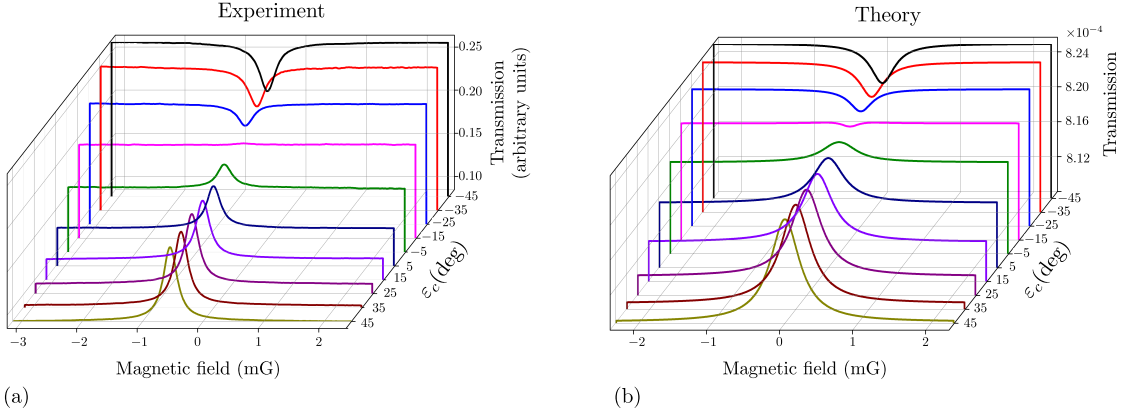}
	\caption{Transformation from EIT to EIA with change in control ellipticity. $ \epsilon_p $ is kept fixed at $+45^\circ$, while $ \epsilon_c $ is varied from $+45^\circ$ to $-45^\circ$. (a) Experimental spectra. (b) Theoretical results; probe Rabi frequency $\Omega_p$ is taken as $0.032$ in units of $\Gamma_e$; control Rabi frequency $\Omega_c$ for different curves is scaled from $\Omega_p$ by the square root of the power ratio as used in the experiment; values of various relaxation rates are as follows: $ \gamma_{t1} = \gamma_{t3} = 0.012\Gamma_e$, $\gamma_{t2} = 0.00003 \Gamma_e $, and $ \gamma_w = 0.000005 \Gamma_e $; transverse B field is taken to have a value of 0.3 mG.}
	\label{MITtoMIA80mW}
\end{figure*}

We next demonstrate that the transformation from EIT to EIA is continuous as the control beam ellipticity is changed from parallel to orthogonal to the probe. Hanle spectra for different values of $\epsilon_c$ varied in steps of $10^\circ$  are shown in Fig.~\ref{MITtoMIA80mW}. The theoretical spectra, shown in part (b) of the figure, closely match the experimental spectra shown in part (a) of the figure. The values of different parameters required for theory are given in the figure caption. The important thing to note is that the theoretical value of $\epsilon_c$ where the transformation occurs is close to the experimental value.

The susceptibility induced in the medium is $\chi= \sum_{g,e}N|\mu|^2 \rho_{eg}/(\hbar \epsilon_0 \Omega^p_{eg})$, where $\mu = 3.797 \times 10^{-29}$ C\,m is the transition dipole matrix element for the D$_2$ line of Cs, $N \approx 10^{10}$ atoms/cc is the number density of Cs atoms at room temperature \cite{RBN17}, $\rho_{eg}$ is the  density matrix element for coherence between ground and excited levels, and $\Omega^p_{eg} $ is the probe Rabi frequency. Since Im$\{\chi\}$ corresponds to absorption inside the medium, we get transmission in the calculated signal by using Im$\{-\chi\}$, along with an arbitrary offset of $10^{-3}$ to make the values positive. The theoretical analysis is done by assuming that the atoms have no velocity distribution. In addition, the intensities in the probe and control fields are assumed to be constant. Both these assumptions imply that the Rabi frequencies necessary to reproduce the experimental results are smaller than those at the center of the two Gaussian beams used for the experimental data, namely 30 \textmu W and 80 \textmu W respectively. This point is described in detail in the work of Renzoni \etal \cite{RCA01}, and discussed in the supplementary file. 

The use of a paraffin-coated vapor cell has two effects. One is that the resonances are much narrower compared to previous work in this field. The second effect, and one that is more important to the work here, is that the transport of atomic coherences with spatially separated beams is only possible in such a cell \cite{ZMK02,XKH08,PCS16}. The inter-beam transport of coherences is due to the fact that the atoms optically polarized by the control beam can enter into the probe beam because of its random motion. During this process, the atom traverses the dark region where it undergoes Larmor precession in the presence of magnetic fields.  In the paraffin coated cell, the coherence induced among the ground states survives for a long time despite countless collisions with cell wall. This is like a Ramsey process, except that the time spent in the laser-free interaction region is random, and the atoms are free to leave and enter either of the beams. This effect has been studied in detail in single-beam nonlinear magneto-optical effect \cite{BYZ98,KIM12}, and is known as the wall-induced Ramsey effect.

\subsection{Effect of control power on the amplitude of the resonances} 
\begin{figure*}[ht]
	\centering
	\includegraphics[width=.9 \textwidth]{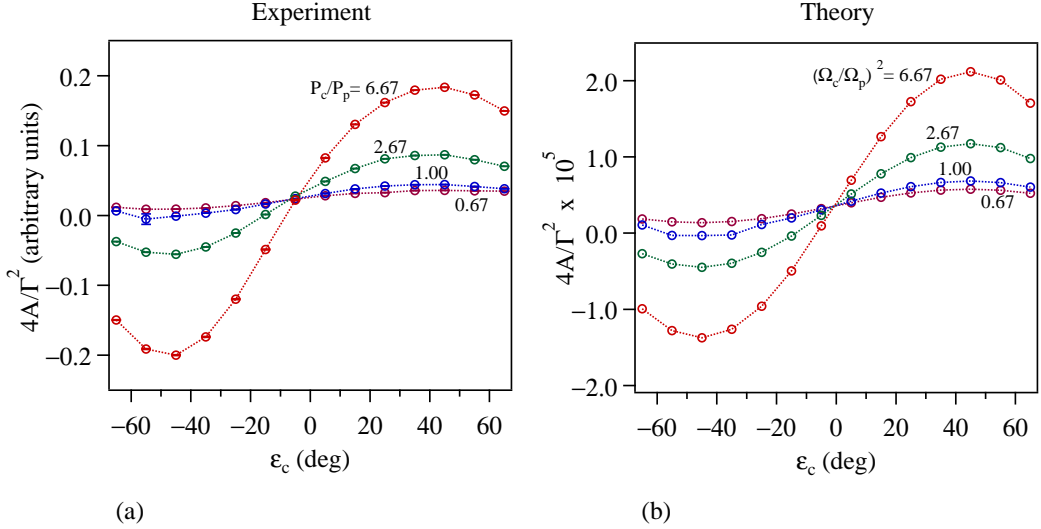}
	\caption{Amplitude of the resonance (as defined in the text) as a function of control ellipticity for 4 values of control power. The probe power is fixed at 30 \textmu W, and the values of control power are given in terms of power ratios. Part (a) shows the experimental results while part (b) shows the theoretical results. Note that the ratio of Rabi frequencies is equal to the square root of the powers. The experimental points have 1$\sigma$ error bars obtained from the fit, but are not seen because they are smaller than the symbol. Theoretical points are obtained from simulations using the same parameters as 
		in Fig.~\ref{MITtoMIA80mW}(b).}
	\label{ampvsell}
\end{figure*}

In this subsection, we discuss the effect of control field power on the amplitude of the Hanle resonances. The resonances for probe transmission are analyzed by fitting the peaks about the line center using a Lorentzian function: 
\begin{equation}
L = \dfrac{A}{(B-B_0)^2+\Gamma^2/4}
\end{equation}
The amplitude of resonance---peak height from the baseline---is given by $4A/\Gamma^2$. The experiments are performed for a fixed probe power of $P_p = 30$ \textmu W. The control power $P_c$ is varied to get the four ratios shown in Fig.~\ref{ampvsell}. The fit gives an error, which is used for the error bar in the experimental data shown in part (a) of the figure. The theoretical results, shown in part (b) of the figure, are obtained with the ratio of Rabi frequencies equal to the square root of the power ratios. The close match between experiment and theory (apart from a scale factor) shows that the theoretical model is quite accurate. This control over the amplitude of narrow resonances implies that the medium can be tuned between sub- and super-luminal light propagation simply by changing the ellipticity of the control beam \cite{PFM00,HHW06,BHN17}. 

\subsection{Zero crossing of the resonance}
\begin{figure}[ht]
	\centering
	\includegraphics[width=.5 \textwidth]{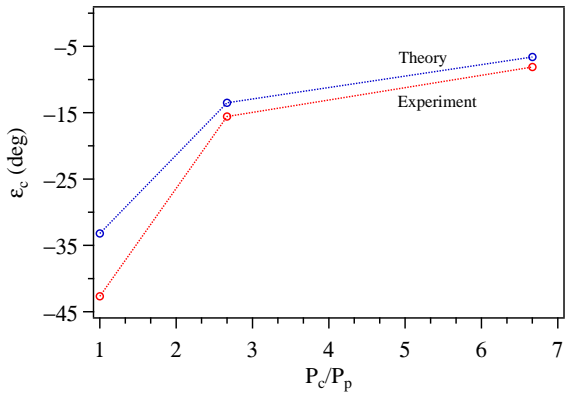}
	\caption{Theoretical and experimental values of the zero point of the Hanle resonance as a function of control power. For ease of comparison to other data, the control power is given as a ratio. }
	\label{zerocrossing}
\end{figure}

The experimental spectra show that the transformation from EIT to EIA, i.e.~where the amplitude of the resonance is zero, does not happen at a value of $ \epsilon_c = 0 $. We have therefore studied the value of $ \epsilon_c $ where the transformation occurs. The results are shown in Fig.~\ref{zerocrossing} for different values of power ratios. The important thing to note is that the theoretical values match the experimental values quite well. The only thing different from the qualitative explanation based on optical pumping (given in the section on theoretical analysis) is when the control beam has a polarization other than circular, because it means that it has both $\sigma^+$ and $\sigma^-$ components. Whenever $ \sigma^- $ transition strength is larger than $ \sigma^+ $ transition (contributed by probe and corresponding component of control), the population gets pumped into the $ \ket{1} $ state. This explanation tells us that for a particular value of $ \epsilon_c $ ($\neq$ 0) the Hanle resonance transforms from EIT to EIA. This also implies that the control power has to be equal or larger for the transformation to occur.

\section{Conclusions}
In summary, we demonstrate a technique of tuning Hanle resonances from EIT to EIA, using spatially separated probe and control beams. The experiments are done using magnetic sublevels of the closed $F_g = 4 \rightarrow F_e = 5 $ transition in the D$_2$ line of $^{133}$Cs. The atoms are contained in a room remperature vapor cell, with paraffin coating on the walls. Such coating is important for atomic coherences to be maintained between the spatially separated beams.

The requirement for EIT is that the two beams have same circular polarizations. The peak transforms to EIA when the polarizations are orthogonal. The transformation occurs only when the control power is much larger than the probe power. We have verified this by studying the transformation for different ratios of powers. The experimental results are explained by a theoretical density-matrix analysis of the sublevels involved in the transition. The model is also able to explain the observed value of control ellipticity where the transformation from EIT to EIA occurs.

The ability to control Hanle resonances from enhanced transmission to enhanced absorption make this a good technique for sensitive magnetometry \cite{GRP16,ASK17} and control over group velocity \cite{PFM00,HHW06}.

\section*{Data Availability}
The raw data used to make Fig.~\ref{ampvsell}(a) is available in the excel file rawdata.xls.

\section*{Acknowledgements}
This work was supported by the Council of Scientific and Industrial Research, India. The authors thank Sumanta Khan, Jay Mangaonkar, and Disha Kapasi for help with the experiments; and S Raghuveer for help with the manuscript preparation.

\section*{Appendix} In this appendix, we discuss the method to approximate the Rabi frequencies for theoretical analysis. The saturation intensity ($I_s$) is not the same for atoms moving with different longitudinal velocities.

The population of excited state after the interaction of atoms with light is given as: 
\begin{equation}
\rho_{22}=\dfrac{\Omega^2/4}{\delta^2+\Omega^2/2+\Gamma^2/4}
\label{rho22}
\end{equation}
where $\delta$ is the detuning of light from resonance, $\Omega$ is the Rabi frequency, and $\Gamma$ is the natural linewidth of excited state. \\For light field on resonance ($\delta=0$) and atoms moving with $v$, Eq.~\eqref{rho22} can be written as:
\begin{equation}
\rho_{22} (v)=\dfrac{\Omega^2/4}{{(k v)}^2+\Omega^2/2+\Gamma^2/4}
\label{rho22v}
\end{equation}
The excited state population by considering Maxwell-Boltzmann (MB) velocity distribution of atoms is
\begin{equation}
\rho_{22} = \int_{-\infty}^{+\infty} \rho_{22}(v) f(v) dv, 
\label{rho22MB}
\end{equation}
where
\begin{equation}
f(v) = \sqrt{\dfrac{m}{2\pi k_B T} } \exp\left(\dfrac{-mv^2}{2 k_B T}\right)
\end{equation}
We solve for the $\Omega/\Gamma$ at saturation by equating Eq.~\eqref{rho22MB} to 1/4. In our theoretical model, we have ignored the longitudinal velocity distribution. Therefore, the effective Rabi frequency for the model assuming atoms with zero longitudinal velocities---calculated by assuming the ratio $ I/I_s $ is same for both model and experiment---is given as 
  \begin{equation}
   (\Omega/\Gamma)_{{\rm model}} = \dfrac{1}{\sqrt{2}}\dfrac{(\Omega/\Gamma)_{{\rm expt}}}{(\Omega/\Gamma)_{I_s, {\rm MB}}}
   \end{equation}

\section*{Author contributions}
VN conceived the study; MB did the experiments;  MB and VB did the theoretical simulation; all authors reviewed the manuscript for intellectual content, and read and approved the final version.

	

	
\end{document}